# Deuteros 2.0: Peptide-level significance testing of data from hydrogen deuterium exchange mass spectrometry

Andy M. Lau[1], Jürgen Claesen[2,3], Kjetil Hansen[1], Argyris Politis[1,*]

[1]Department of Chemistry, King's College London, 7 Trinity Street, London, SE1 1DB, United Kingdom, [2]Microbiology Unit, SCK•CEN, Mol, 2600, Belgium, [3]I-Biostat, Data Science Institute, Hasselt University, Hasselt, 3500, Belgium

*To whom correspondence should be addressed.

## Abstract

**Summary:** Hydrogen deuterium exchange mass spectrometry (HDX-MS) is becoming increasing routine for monitoring changes in the structural dynamics of proteins. Differential HDX-MS allows comparison of individual protein states, such as in the absence or presence of a ligand. This can be used to attribute changes in conformation to binding events, allowing the mapping of entire conformational networks. As such, the number of necessary cross-state comparisons quickly increases as additional states are introduced to the system of study. There are currently very few software packages available that offer quick and informative comparison of HDX-MS datasets and even fewer which offer statistical analysis and advanced visualization. Following the feedback from our original software Deuteros, we present Deuteros 2.0 which has been redesigned from the ground up to fulfil a greater role in the HDX-MS analysis pipeline. Deuteros 2.0 features a repertoire of facilities for back exchange correction, data summarization, peptide-level statistical analysis and advanced data plotting features.

**Availability:** Deuteros 2.0 can be downloaded from https://github.com/andymlau/Deuteros_2.0 under the Apache 2.0 license. Installation of Deuteros 2.0 requires the MATLAB Runtime Library available free of charge from MathWorks (https://www.mathworks.com/products/compiler/matlab-runtime.html) and is available for both Windows and Mac operating systems.

**Contact:** argyris.politis@kcl.ac.uk

## 1 Introduction

Hydrogen-deuterium exchange mass spectrometry (HDX-MS) is a structural MS technique which can be used to monitor changes in both protein structure and conformation (Marcsisin and Engen, 2010). Labile amide hydrogens of the protein backbone exchange with deuterium in the surrounding environment. Exchange competence depends on both structural and physiochemical factors (Konermann, et al., 2011). HDX leads to increases in the protein mass which can be detected by high resolution MS. Proteolytic digestion allows further localization of HDX to the peptide-level (Marcsisin and Engen, 2010).

HDX-MS experiments are commonly performed in a differential manner. Differential HDX-MS envisions that two or more datasets collected from the same protein but in different conditions are compared with one another to study changes in regional deuterium uptake as a function of environmental changes. Such changes commonly include the addition of external interacting species such as substrates and inhibitors (Ahdash, et al., 2019; Chalmers, et al., 2011), environmental changes such as to pH and membrane lipid composition (Li, et al., 2014; Martens, et al., 2018; Murphy, et al., 2019) or even modifications to the target protein itself such as mutations and other covalent customizations (Martens, et al., 2018).

Raw data acquired from HDX-MS must undergo several processing steps before it can be interpreted (Claesen and Burzykowski, 2017; Wales, et al., 2013). The first of these steps requires the identification of peptide ions from the raw spectra and fragmentation data. Next, the isotopic distribution of each identified peptide ion is processed to determine its corresponding mass. The mass of a non-deuterated reference peptide is subtracted from its deuterated equivalent to determine the level of deuterium uptake. Finally, the deuterium uptake of each peptide is used to develop a spatial representation of the protein conformation. For differential HDX-MS, data from a reference dataset (e.g. wild-type protein) is subtracted from data belonging to an altered state (e.g. mutant protein). The significance of the changes identified from differential analyses can be evaluated using a range of statistical models including global threshold tests (Houde, et al., 2011), mixed effects models (Hourdel, et al., 2016) and hybrid significance tests (Hageman and Weis, 2019). While the first two steps of peptide identification and mass determination can be well managed using existing software such as ProteinLynx Global Server and DynamX (Waters Corporation), how the final step of statistical analysis and visualization is best performed is area of active discussion.

In 2019, we released our software Deuteros (Lau, et al., 2019), designed to perform rapid analysis and visualization of data from the Waters HDX-MS platform. Parallel to developments in the HDX-MS field, we present here Deuteros 2.0 which has been redesigned from the ground up with new features intended to further streamline analysis of HDX-MS data. In summary, Deuteros 2.0 contains several notable features including automatic back-exchange correction, data summary, statistical analysis (including a new peptide-level significance test), and advanced visualization features such as volcano plots and exporting results to molecular visualization.

## 2 How does it work?

### 3.1 Overview of Deuteros 2.0

Several notable improvements have been made during the redesign of Deuteros 2.0. Firstly, the input data for Deuteros 2.0 takes the form of the DynamX cluster file. The cluster file contains a replicate-level breakdown of an HDX-MS dataset with the following hierarchy (highest level to lowest level): protein, state, peptide, timepoint, charge state and technical replicates. An unlimited number of protein states can be included in a single cluster file, allowing greater portability and accountability of data. Users can optionally select automatic back-exchange correction of their data prior to downstream processing. To do this, an experimental state corresponding to back-exchange control data can be included in the cluster file and selected during data import.

The data imported into Deuteros 2.0 undergoes several transformations: 1) calculation of intensity-weighted average deuterium uptake and standard deviation - for each peptide, timepoint and state. An additional step removes data points belonging to redundant and/or poor quality charge states; 2) identification of common peptides between two user-selected states A and B (e.g. holo vs apo); 3) calculation of deuterium uptake differences for each peptide $i$, as $\Delta DU_i = DU_{B,i} - DU_{A,i}$; 4) treatment of the uptake difference data with a user-selected statistical model to identify regions of the target protein exhibiting state-specific uptake patterns. For determining whether differences in deuterium uptake are statistically significant, Deuteros 2.0 offers a choice of two statistical models (Fig. 1): a peptide-level significance test and a recently published hybrid significance test (Hageman and Weis, 2019).

To aid data interpretation, we have expanded the interface of Deuteros 2.0 with statistics-enabled facilities for evaluating the kinetics plot of individual peptides. For differential HDX-MS, we have included several graphical data presentation formats including Woods, volcano and barcode plots. For visualisation on structural models, Deuteros 2.0 can now output statistically evaluated results to both PyMOL and Chimera. Deuteros 2.0 is also compliant with community guidelines on HDX-MS data reporting by automatically producing a summary of data quality for the imported dataset (Masson, et al., 2019).

### 3.2 Peptide-level significance testing

The peptide-level significance test uses multiple regression to fit deuterium uptake kinetics to a linear model (eq. 1).

$$D_{ij} = \beta_0 + \beta_s s_i + \beta_t t_j + \beta_{st}(s_i t_j) + \varepsilon_{ij} \qquad (1)$$

Where $D_{ij}$ is the deuterium content of a peptide from state $s_i$, at labelling time $t_j$ and residual $\varepsilon_{ij} \sim N(0, \sigma_{ij}^2)$. The linear model tests if changes in the deuterium content of the peptide are associated with changes in the protein state ($\beta_s$), the deuteration time point ($\beta_t$), or both ($\beta_{st}$). Additionally, the probability of detecting false positives is controlled by the Benjamini-Hochberg procedure for multiple testing (Benjamini and Hochberg, 1995). This False Discovery Rate controlling method adjusts the p-value of each test using $m$, the number of hypotheses tested, and $i$ the ordered rank of the *p*-value (eq. 2):

$$p^* = p \times (m/i) \qquad (2)$$

The null hypothesis is rejected on the basis that the adjusted *p*-value is less than the significance level.

To provide users with flexibility with comparing different statistical models, we have included an additional statistical test, namely the hybrid significance test (Hageman and Weis, 2019). Briefly, the hybrid significance test is a two-pronged statistical test which first evaluates whether the deuterium uptake difference of a peptide between two states is greater than a threshold value calculated to a user-defined significance level. A Welch's t-test is then employed to confirm the significance (Hageman and Weis, 2019).

### 3.3 Data Visualization

Deuteros 2.0 improves on the original Deuteros software by including several redesigned and new data visualization methods. These include improvements to the 'Coverage Plot' section of the software through the addition of comparative facilities allowing users to quickly determine changes in coverage and redundancy between experiments. We anticipate that this feature will extend the usefulness of Deuteros 2.0 in aiding users with making experimental design choices, such as evaluating whether or not a change in digestion enzyme improves coverage to certain regions of targets. We foresee that this feature may be particularly useful for studies on difficult targets such as membrane proteins. Coverage can also be assessed using a new barcode plot which includes the level of deuterium uptake over the full length of the protein for each experimental timepoint and for each state.

Under the 'Advanced Plot' section, users can assess data through one of the statistics-enabled plots. These include the Woods and volcano plots. In Woods plots, each peptide is visualized as a single horizontal bar along a residue number axis with the bar length representing the length of the peptide and the bar height representing the magnitude of change in deuterium uptake between two selected states. Woods plots are particularly useful for data presentation as they additionally depict both coverage and redundancy. The volcano plot is only enabled for datasets assessed using the hybrid significance test. The volcano plot format visualizes deuterium uptake differences along the horizontal axis and statistical significance on the vertical axis. In both Woods and volcano formats, datapoints are subdivided into subplots for each experimental timepoint. Peptides with significant positive changes in deuterium uptake (areas deprotected upon change), significant negative changes (areas protected upon change) and non-significant changes follow a red, blue and grey color scheme respectively. All figures produced from the 'Advanced Plot' section have been enabled with interactivity – users can select individual datapoints to spawn 'datatips' with additional information about the peptide of interest.

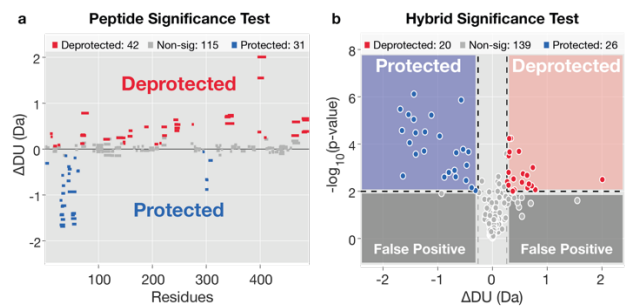

**Fig. 1. Statistical testing in Deuteros 2.0.** Two statistical tests are offered in Deuteros 2.0: (a) peptide and (b) hybrid significance tests. Statistical models are applied to differential HDX-MS data to identify peptides which show statistically significant changes in deuteration between experimental states. Significant peptides are described as regions exhibiting deprotection or protection according to the direction of the deuteration difference. Data shown represents the XylE Δ(E397Q-WT) dataset from (Martens, et al., 2018).

Data processed and statistically treated using Deuteros 2.0 can be further visualized on molecular structures. Using this feature, protein data bank (PDB) files in either PyMOL (Schrödinger, 2015) or Chimera (Pettersen, et al., 2004) can be formatted to highlight regions of the protein identified through Deuteros 2.0 to be statistically significant. Coverage, redundancy and significant peptide regions determined through Woods or volcano plots are enabled for exporting to molecular graphics.

## Acknowledgements

We would like to thank Justin Benesch, Angela Gehrckens (University of Oxford) and Ruyu Jia (King's College London) for assistance with software testing and feedback.

## Funding

This work was supported by grants from the London Interdisciplinary Biosciences Consortium (LIDo) BBSRC Doctoral Training Partnership (BB/M009513/1) and the Leverhulme Trust (RPG-2019-178).

*Conflict of Interest:* none declared.

## References

Ahdash, Z., *et al.* HDX-MS reveals nucleotide-dependent, anti-correlated opening and closure of SecA and SecY channels of the bacterial translocon. *Elife* 2019;8.

Benjamini, Y. and Hochberg, Y. Controlling the False Discovery Rate: A Practical and Powerful Approach to Multiple Testing. *Journal of the Royal Statistical Society: Series B (Methodological)* 1995;57(1):289-300.

Chalmers, M.J., *et al.* Differential hydrogen/deuterium exchange mass spectrometry analysis of protein-ligand interactions. *Expert Rev Proteomics* 2011;8(1):43-59.

Claesen, J. and Burzykowski, T. Computational methods and challenges in hydrogen/deuterium exchange mass spectrometry. *Mass Spectrom Rev* 2017;36(5):649-667.

Hageman, T.S. and Weis, D.D. Reliable Identification of Significant Differences in Differential Hydrogen Exchange-Mass Spectrometry Measurements Using a Hybrid Significance Testing Approach. *Anal Chem* 2019;91(13):8008-8016.

Houde, D., Berkowitz, S.A. and Engen, J.R. The utility of hydrogen/deuterium exchange mass spectrometry in biopharmaceutical comparability studies. *J Pharm Sci* 2011;100(6):2071-2086.

Hourdel, V., *et al.* MEMHDX: an interactive tool to expedite the statistical validation and visualization of large HDX-MS datasets. *Bioinformatics* 2016;32(22):3413-3419.

Konermann, L., Pan, J. and Liu, Y.H. Hydrogen exchange mass spectrometry for studying protein structure and dynamics. *Chem Soc Rev* 2011;40(3):1224-1234.

Lau, A.M.C., *et al.* Deuteros: software for rapid analysis and visualization of data from differential hydrogen deuterium exchange-mass spectrometry. *Bioinformatics* 2019;35(17):3171-3173.

Li, J., *et al.* Hydrogen-deuterium exchange and mass spectrometry reveal the pH-dependent conformational changes of diphtheria toxin T domain. *Biochemistry* 2014;53(43):6849-6856.

Marcsisin, S.R. and Engen, J.R. Hydrogen exchange mass spectrometry: what is it and what can it tell us? *Anal Bioanal Chem* 2010;397(3):967-972.

Martens, C., *et al.* Direct protein-lipid interactions shape the conformational landscape of secondary transporters. *Nat Commun* 2018;9(1):4151.

Masson, G.R., *et al.* Recommendations for performing, interpreting and reporting hydrogen deuterium exchange mass spectrometry (HDX-MS) experiments. *Nat Methods* 2019;16(7):595-602.

Murphy, R.E., *et al.* Structural and biophysical characterizations of HIV-1 matrix trimer binding to lipid nanodiscs shed light on virus assembly. *J Biol Chem* 2019;294(49):18600-18612.

Pettersen, E.F., *et al.* UCSF Chimera--a visualization system for exploratory research and analysis. *J Comput Chem* 2004;25(13):1605-1612.

Schrödinger, L. The PyMOL Molecular Graphics System, Version 1.8. 2015.

Wales, T.E., Eggertson, M.J. and Engen, J.R. Considerations in the analysis of hydrogen exchange mass spectrometry data. *Methods Mol Biol* 2013;1007:263-288.